\documentclass[sigconf]{acmart}
\usepackage{amsmath}
\usepackage{algorithmic}
\usepackage{subcaption}
\usepackage{caption}
\usepackage{array}
\usepackage{makecell}
\newcommand{\etal}{\textit{et al.}}
\newcommand{\ie}{\textit{i.e.}}

\AtBeginDocument{%
  \providecommand\BibTeX{{%
    \normalfont B\kern-0.5em{\scshape i\kern-0.25em b}\kern-0.8em\TeX}}}
 
\copyrightyear{2025}
\acmYear{2025}
\setcopyright{cc}
\setcctype{by}
\acmConference[GLSVLSI '25]{Great Lakes Symposium on VLSI 2025}{June 30-July 2, 2025}{New Orleans, LA, USA}
\acmBooktitle{Great Lakes Symposium on VLSI 2025 (GLSVLSI '25), June 30-July 2, 2025, New Orleans, LA, USA}
\acmDOI{10.1145/3716368.3735290}
\acmISBN{979-8-4007-1496-2/2025/06}
 
\begin{document}

\title{Quantum Transfer Learning to Boost Dementia Detection}

\author{Sounak Bhowmik}
\affiliation{%
  \institution{University of Tennessee, Knoxville}
  \city{Tennessee}
  \country{USA}}
\email{sbhowmi2@vols.utk.edu}

\author{Talita Perciano}
\affiliation{%
  \institution{Lawrence Berkeley National Laboratory}
  \city{California}
  \country{USA}}
\email{tperciano@lbl.gov}

\author{Himanshu Thapliyal}
\affiliation{%
  \institution{University of Tennessee, Knoxville}
  \city{Tennessee}
  \country{USA}}
\email{hthapliyal@utk.edu}
\renewcommand{\shortauthors}{Sounak, Talita and Himanshu \etal}

\begin{abstract}
Dementia is a devastating condition with profound implications for individuals, families, and healthcare systems. Early and accurate detection of dementia is critical for timely intervention and improved patient outcomes. While classical machine learning and deep learning approaches have been explored extensively for dementia prediction, these solutions often struggle with high-dimensional biomedical data and large-scale datasets, quickly reaching computational and performance limitations. To address this challenge, quantum machine learning (QML) has emerged as a promising paradigm, offering faster training and advanced pattern recognition capabilities. This work aims to demonstrate the potential of quantum transfer learning (QTL) to enhance the performance of a weak classical deep learning model applied to a binary classification task for dementia detection. Besides, we show the effect of noise on the QTL-based approach, investigating the reliability and robustness of this method. Using the OASIS 2 dataset, we show how quantum techniques can transform a suboptimal classical model into a more effective solution for biomedical image classification, highlighting their potential impact on advancing healthcare technology.
\end{abstract}



\maketitle

\section{Introduction}
Dementia is a neurodegenerative disorder causing a decline in cognitive and social abilities, affecting a large number of people around the world. Dementia affects the patient's memory, thinking, and perception, interfering with their daily activities and social lives. Early and accurate detection of dementia is essential to slow down the progression of the disease and improve the patients' lives. Traditional diagnostic methods, including clinical trials, magnetic resonance imaging (MRI), and positron emission tomography (PET), face challenges such as sensitivity, specificity, and manual error. Recent advancements in machine learning \cite{wang2024understanding, pateria2024comprehensive} have shown promising results, enhancing detection accuracy and identifying complex patterns in biomedical data. However, classical machine learning approaches have computational limits while processing high-dimensional neuroimaging datasets \cite{patil2024machine}. 

Quantum computing has shifted the computational paradigm by leveraging quantum mechanical phenomena like superposition and entanglement to perform computations intractable by classical methods. Quantum machine learning (QML) combines quantum computational abilities with powerful machine learning algorithms. Especially quantum transfer learning (QTL)~\cite{bhowmik2024transfer} uses a combination of feature-extraction techniques provided by classical machine learning paradigms and quantum neural networks (QNN), accelerating computations and excelling in specific tasks. QTL uses pre-trained classical models to extract insightful features and processes them through quantum neural networks to classify data. This hybrid strategy utilises the full potential of established classical methods while capitalising on the computational benefits of a quantum processor.

%==========================
\begin{figure}[ht]
    \centering

    % First row: Demented
    \begin{minipage}{0.15\columnwidth} % Adjusted for the label
        \centering
        \rotatebox{90}{\textbf{Demented}} % Bold label for better visibility
    \end{minipage}%
    \hfill
    \begin{minipage}{0.27\columnwidth}
        \centering
        \includegraphics[width=\linewidth]{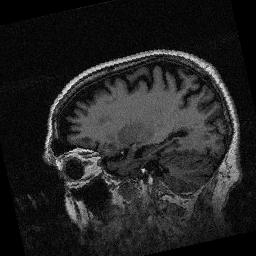}
    \end{minipage}%
    \hfill
    \begin{minipage}{0.27\columnwidth}
        \centering
        \includegraphics[width=\linewidth]{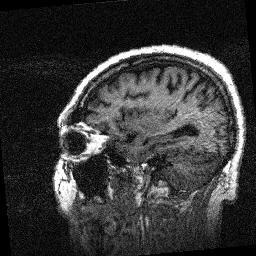}
    \end{minipage}%
    \hfill
    \begin{minipage}{0.27\columnwidth}
        \centering
        \includegraphics[width=\linewidth]{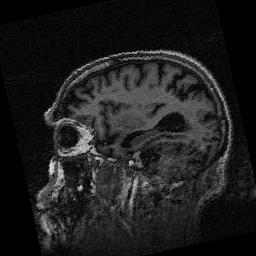}
    \end{minipage}

    \vspace{1em} % Space between rows

    % Second row: Non-Demented
    \begin{minipage}{0.15\columnwidth} % Adjusted for the label
        \centering
        \rotatebox{90}{\textbf{Non-demented}} % Bold label for better visibility
    \end{minipage}%
    \hfill
    \begin{minipage}{0.27\columnwidth}
        \centering
        \includegraphics[width=\linewidth]{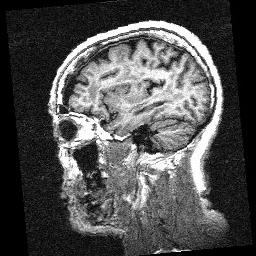}
    \end{minipage}%
    \hfill
    \begin{minipage}{0.27\columnwidth}
        \centering
        \includegraphics[width=\linewidth]{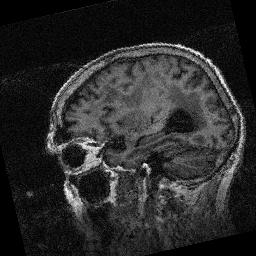}
    \end{minipage}%
    \hfill
    \begin{minipage}{0.27\columnwidth}
        \centering
        \includegraphics[width=\linewidth]{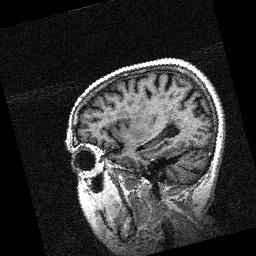}
    \end{minipage}
    \caption{Samples from the `demented' and `non-demented' classes in the OASIS 2 dataset.}
    \label{Fig: Datasamples}

\end{figure}

%==========================================

We can find some interesting works to detect Alzheimer's Disease and Related Dementias (ADRD) using quantum machine learning in the literature. Abebech~\etal~proposed an ensemble method based on quantum machine learning classifiers to detect Alzheimer's disease in~\cite{jenber2024deep}. They used powerful classical feature extractors such as VGG16 and ResNet50. They used the extracted features in a quantum support vector machine classifier to classify brain MRI images as non-demented, mild, very mild, and moderately demented. Giacomo~\etal~\cite{cappiello2024quantum} have compared the performance between classical and quantum kernel methods while predicting Alzheimer's disease based on the DARWIN dataset of handwriting samples. They observed that quantum kernels sometimes outperform classical ones, keeping performance consistent with that of later ones. Moona~\etal~\cite{mazher2024hybrid} proposed a hybrid quantum-classical convolutional neural network model for Alzheimer's disease classification. Their model was reliable and robust because of its optimal performance over various datasets. Ryan~\etal~\cite{kim2023hybrid} proposed a hybrid quantum-classical machine learning model for dementia detection. Introducing quantum features could significantly improve the performance of the classical model. Kuan-Cheng~\etal~\cite{chen2024compressedmediqhybridquantummachine} introduced a novel quantum-classical pipeline to handle the difficulties in processing high-dimensional neuroimaging data. This pipeline comprises classical HPC nodes to pre-process high-dimensional data, followed by classical convolutional neural networks to extract features and feed the extracted high-quality features to the quantum support vector machine classifier for further analysis.

In this work, we aim to show how to use quantum transfer learning to enhance a weak classifier effectively. Also, it investigates the practical utility of this approach in a noisy environment. We explored one such application to predict dementia using neuroimaging biomedical data. Starting with training a suboptimal baseline classical convolutional neural network (CNN), we applied quantum transfer learning to build a resilient hybrid quantum-classical neural network model, which showed consistent and reliable performance for the selected task.

Our contributions are as follows:
\begin{itemize}
    \item We develop a pipeline that leverages quantum transfer learning to train hybrid quantum-classical neural network models dedicated to dementia prediction.
    \item We demonstrate that applying QTL can substantially improve the performance of a weak classical model, as confirmed by comparisons with various QTL-based hybrid model configurations.
    \item We analyse the impact of noise on QTL-based hybrid models to assess their resilience in noisy intermediate-scale quantum (NISQ) devices.
\end{itemize}
The rest of the paper is organised as follows: Section~\ref{sec: background} presents some preliminary ideas on quantum transfer learning (QTL). Section~\ref{sec: method} discusses the dataset, the architecture of the QTL-based hybrid models, and the experimental setup. Following up with Section~\ref{sec: results}, we present the results obtained in the experiment. At last, we conclude the work in Section~\ref{sec: conclusion}.

\section{Background}\label{sec: background}
Theoretically, we can formulate quantum transfer learning (QTL) as a composition of classical and quantum mappings acting on high-dimensional classical data. For example, $ f_{\mathrm{cl}}: \mathbb{R}^{M} \to \mathbb{R}^{D} $ is a pre-trained classical deep learning model, initially trained on a large, generic dataset to extract domain-agnostic features. Given a data instance \( x \in \mathbb{R}^{M} \) (\textit{e.g.}, a pre-processed MRI image for dementia diagnosis), the classical model outputs a latent vector \( z = f_{\mathrm{cl}}(x) \in \mathbb{R}^{D} \). Next, we introduce a parameterized encoding map \( U_{\mathrm{enc}}(z) \) that converts \( z \) into an \( n \)-qubit quantum state \(|\psi(z)\rangle = U_{\mathrm{enc}}(z)|0\rangle^{\otimes n} \in \mathcal{H} \), to transition into the quantum realm~\cite{weigold2020data}. Here, \(\mathcal{H}\) is a Hilbert space of dimension \(2^n\) and \( n = O(\log D) \). This encoding is designed to preserve geometric relationships in \(\mathbb{R}^{D}\) or to meaningfully transform them into the amplitudes of \(|\psi(z)\rangle\). A subsequent parameterised quantum circuit (PQC) \( U(\theta) \) acts on this state, producing \(|\phi_{\theta}(z)\rangle = U(\theta)|\psi(z)\rangle\). Then a suitable positive operator-valued measure (POVM) \(\{M_y\}\) corresponding to class labels \(y\), performs the classification, and the predicted probability distribution emerges as,
\[
p(y|z;\theta) = \langle \phi_{\theta}(z)|M_y|\phi_{\theta}(z)\rangle.
\]
Training the quantum circuit involves adjusting \(\theta\) to minimize a task-dependent empirical risk, for instance,
\[
\min_{\theta} \frac{1}{N}\sum_{i=1}^N L(p(y_i|z_i;\theta), y_i),
\]
where \( L \) could be a cross-entropy loss function and \(\{(z_i,y_i)\}\) represents the transformed training samples and labels.

This hybrid formalism taps into the representative capacity of quantum states to enrich decision boundaries, potentially achieving class separability beyond classical limitations. Unlike classical neural networks, which realize functions \( f: \mathbb{R}^{D} \to \mathbb{R}^{K} \) (for \( K \) classes) through nested compositions of linear maps and nonlinear activations, QTL leverages the exponential dimensionality and unitary operations of quantum circuits. By encoding a classically derived latent vector \( z \) into \(|\psi(z)\rangle\), the quantum model can implement transformations that correspond to complex, high-order feature interactions. These transformations may be less susceptible to classical local minima or require fewer parameters to represent certain hypothesis classes. Theoretical works suggest that certain distributions or label functions that are difficult to approximate classically might admit more compact representations in a quantum Hilbert space. Thus, QTL can be framed as solving a joint optimisation problem over \(\theta\) that blends the stability and symbolic power of a pre-trained classical front-end with the advanced function approximation capabilities of variational quantum circuits, potentially yielding improved predictive performance for intricate tasks like dementia prediction.

\section{Methodology}\label{sec: method}
We begin this experiment by training a classical baseline convolutional network. We assess the overall performance by 4-fold cross-validation, training on the entire dataset, and validating on a separate test dataset. In the next step, we apply different configurations of quantum and classical transfer learning methods, only to observe if fine-tuning the transfer learning models can improve the performance of the baseline classical model. 

In the following sections, we shall dive deeper into the dataset, the architectural details of the baseline and the transfer learning models, and the experimental setup.

%==============================================

\begin{figure}[htbp]
    \centering
    \includegraphics[width=0.85\linewidth]{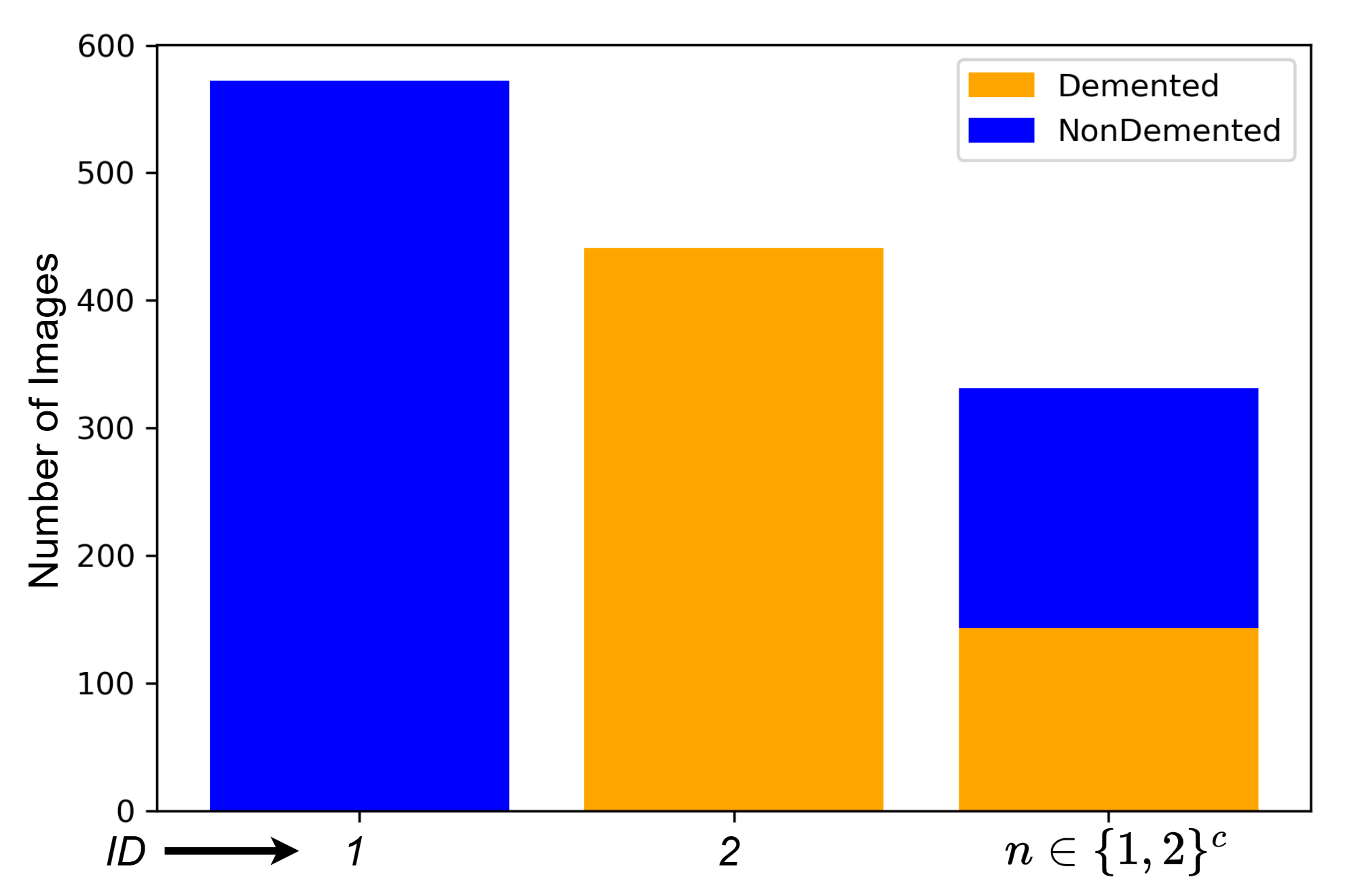}
    \caption{Data distribution based on patient \textit{IDs}.}
    \label{fig:data_dist}
\end{figure}

%==============================================

\subsection{Dataset}

In this experiment, we have used a preprocessed version of the OASIS-2 dataset \cite{OASIS_preprocessed, waldo2024dementia}, consisting of brain MRI images of 72 non-demented and 64 demented subjects, within 60 to 96 years of age. The dataset comprises 584 \textit{`demented'} and 760 \textit{`non-demented'} samples. Each patient is distinguished by a unique \textit{ID}. Figure~\ref{fig:data_dist} shows the distribution of the number of images across different subjects and classes in the dataset. Here we can see that a large number of non-demented (the first blue column) images are associated with \textit{ID}=1. Similarly, a significantly large number of demented (the second yellow column) images are associated with \textit{ID}=2. The brain images from other patients with \textit{IDs} apart from \(\{1,2\}\) have almost evenly distributed demented and non-demented images, shown in the third column. In such a scenario, to prevent data leakage and construct a balanced train and test dataset, we put all the images from \textit{IDs} = 1 and 2 into the train set. We chose 30\% randomly from the rest of the subjects and built the test dataset with all the brain images associated with them. The remaining 70\% were additionally taken into the train set. 

We resized the images to a fixed dimension of (128, 128) and converted them to grayscale for this experiment to reduce the computational overhead.

\begin{figure}[htbp]
    \centering
    \includegraphics[width=0.85\linewidth]{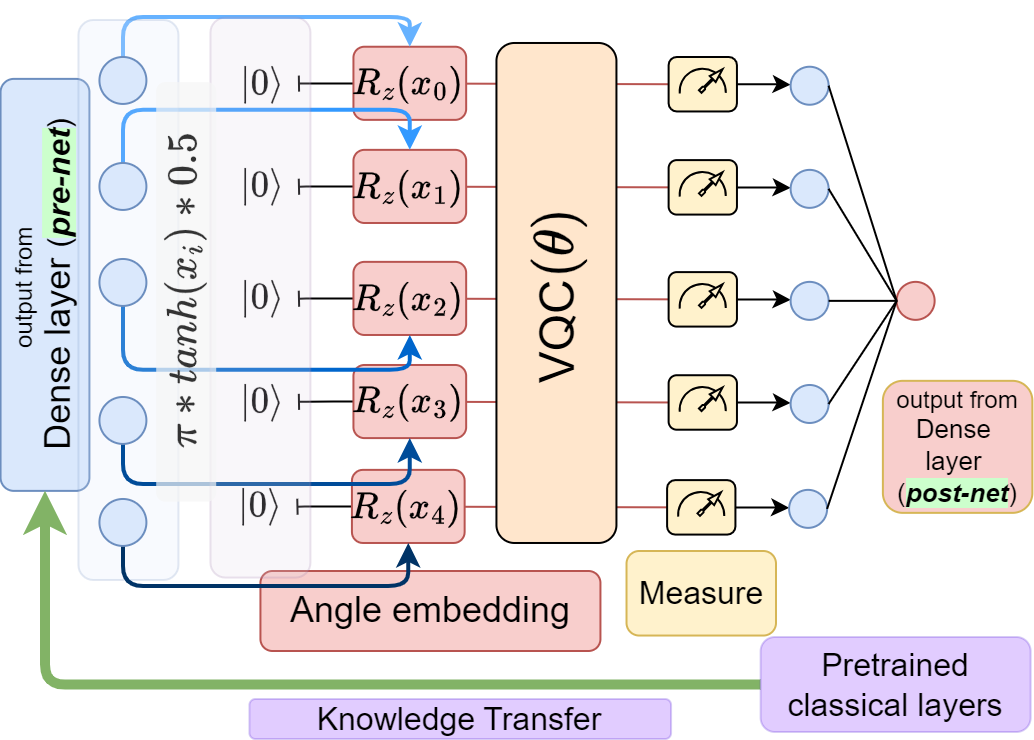}
    \caption{The setup to apply quantum transfer learning to classical data.}
    \label{fig:workflow_tl}
\end{figure}

%==============================================
\begin{table}[h!]
    
    \centering
    \caption{Architecture of the best-performing Classical Convolutional Neural Network Model}
    \label{tab:classical_model}
    \begin{tabular}{ccc}
        \hline
        \textbf{Layer} & \textbf{Hyper parameters} & \textbf{Activation} \\ \hline
        Convolution 2D & \makecell[l]{i/p channel: 1 \quad o/p channel: 8\\ kernel: 4x4 \quad \quad stride: 2} & ReLU \\ \hline
        Maxpool 2D & kernel: 2x2 \quad \quad stride: 1 & N.A. \\ \hline
        
        Convolution 2D & \makecell[l]{i/p channel: 8 \quad o/p channel: 16\\ kernel: 8x8 \quad \quad stride: 2} & ReLU \\ \hline
        Maxpool 2D & kernel: 2x2 \quad \quad stride: 1 & N.A. \\ \hline

        Convolution 2D & \makecell[l]{i/p channel: 16 \quad o/p channel: 32\\ kernel: 8x8 \quad \quad stride: 2} & ReLU \\ \hline
        Maxpool 2D & kernel: 2x2 \quad \quad stride: 1 & N.A. \\ \hline
        
        Convolution 2D & \makecell[l]{i/p channel: 32 \quad o/p channel: 64\\ kernel: 4x4 \quad \quad stride: 1} & ReLU \\ \hline

        Dense & units: 5 \quad \quad  dropout: 0.5 & ReLU\\ \hline
        Dense & units: 1  & sigmoid\\ 
        \hline
        \bottomrule
    \end{tabular}

\end{table}
%==============================================

\subsection{Baseline Model}
The baseline classical convolutional model follows a LeNet-based architecture. It has an initial set of convolutional layers, \ie, the feature extractors, followed by dense layers. The architectural details are presented in table~\ref{tab:classical_model}. The baseline model represents a weak classifier, producing suboptimal results. We aim to use transfer learning to enhance its performance without significantly increasing the computational burden. 

%==============================================
\begin{figure}[htbp]
    \centering
    \includegraphics[width=0.85\linewidth]{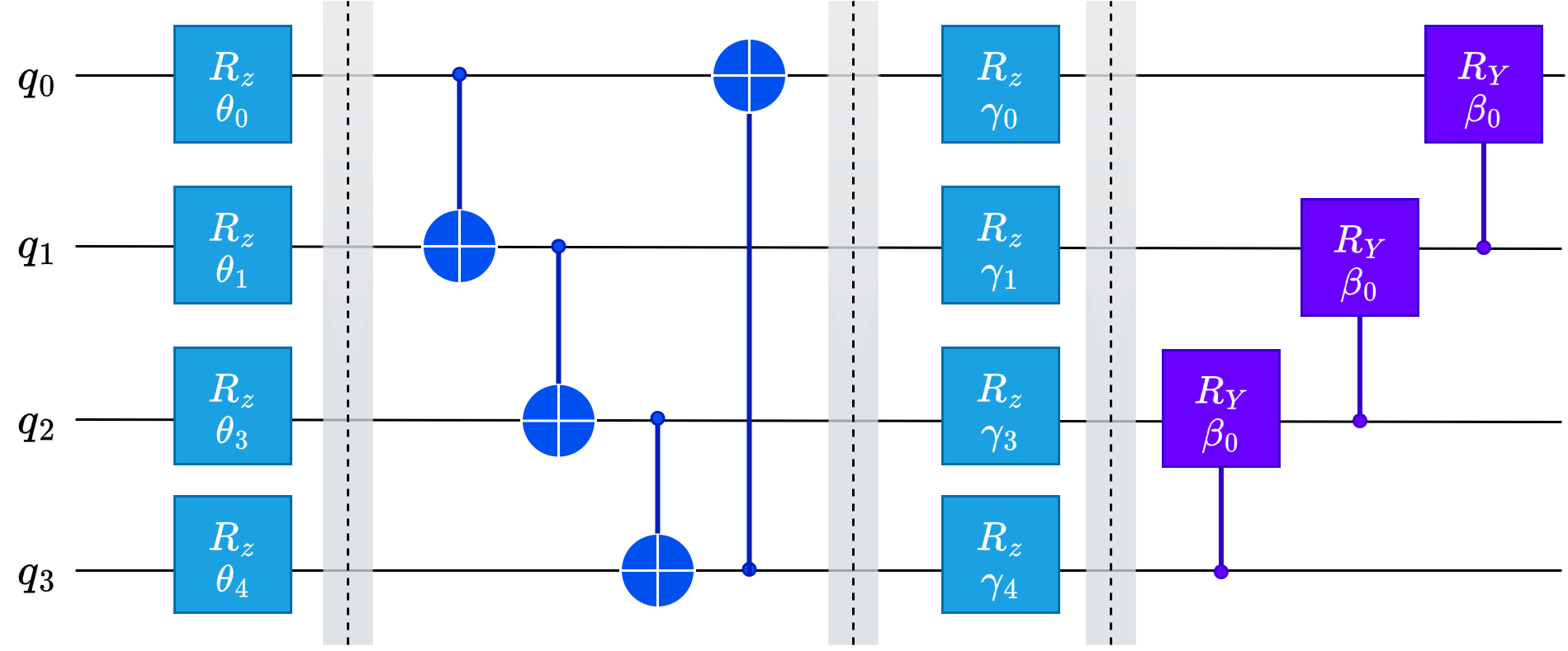}
    \caption{Ansatz comprises parameterised Rz gates, entangling C-NOT ring and controlled, parameterised Ry gates.}
    \label{fig:ansatzA}
\end{figure}

%==============================================

\subsection{Quantum Transfer Learning (QTL) Models}
This section will discuss the architectures of different transfer learning models used in this experiment. Before that, we shall explore the pipeline to apply quantum transfer learning to a classical model, depicted in Figure~\ref{fig:workflow_tl}. 

We begin by freezing the initial convolutional layers in the pre-trained baseline classical model. Then we replace the final dense layers of the model with a dressed quantum network or DQN. A DQN comprises four essential parts. The first dense layer, called a pre-net, maps the input feature dimension to the number of qubits in the following variational quantum circuit (VQC). This layer removes the dependency of the width of the quantum circuit on the input data dimension. 

We apply the \textit{tanh} activation function at the output of pre-net to map them within a range of [1, -1], followed by a multiplication by $\frac{\pi}{2}$ to map the data within a range of $[\frac{\pi}{2}, -\frac{\pi}{2}]$, which is suitable for angle embedding. We used angle embedding in this experiment as we found this method to be the most stable and least sensitive to data quality. 

The angle embedding prepares the quantum state from the classical input data. The VQC then processes this quantum state. The VQC consists of several parameterised rotational gates and entangling circuits. We measure the outputs of the VQC to obtain classical expectation values, which are further processed by the post-net, another classical dense layer, which produces raw logit values, used in classification. The output dimension of the post-net exactly matches the number of classes in a multiclass classification. We use only one neuron at the output, as this work deals with binary classification.

We used two different ansatzes in the VQC of the DQN in the quantum transfer learning (QTL) setups. The four-qubit versions of these circuits are depicted in Figure~\ref{fig:ansatzA}. We constructed different variations of the Ansatz by varying the number of qubits from 3 to 10 and the number of repetitions from 2 to 4 in different transfer learning models.

\subsection{Classical Fine Tuning}
To show the superiority of the QTL-based models over classical transfer learning, we fine-tuned the baseline classical model by freezing the pre-trained convolutional layers and fine-tuning the dense layers in a similar setup. These classical models are Glorot initialised (only the dense layers, remember that we freeze the convolutional layers) several times and trained on the entire dataset to ensure that the reason for an overall suboptimal performance is not local optima. In the results, we have reported the best-performing model among them.

\subsection{Experimental Setup for QTL Training}
This experiment used a binary classification problem to predict dementia based on brain MRI images with the help of quantum transfer learning. We first trained a classical baseline model, which produced suboptimal results, and applied quantum transfer learning. We have cross-validated and trained the QTL-based models derived from the baseline. We also trained classical transfer learning based equivalent classical models to compare their performances.

The training environment was written using the PyTorch library. We used PennyLane's \textit{lightning.qubit} simulator with the differentiation method \textit{`adjoint'} to train the quantum circuits. Also, we used Pennylane's \textit{default.qubit} and Qiskit's \textit{Aer backend} to run noiseless inferences. 

We have used Ionq's 36-qubit Forte-1 noisy simulator to get inferences from the QTL-based hybrid models on the test dataset. The simulator has a detailed gate-level characterisation of the potential noise sources in Ionq's Forte-1 hardware. The noise model is approximated by using a depolarisation error channel.

To optimise all the models, we used the Adam optimiser with a learning rate of $10^{-4}$, and a step learning rate scheduler that reduces the learning rate to 25\% every 10 steps to train all the models. We used the binary cross-entropy loss function as it suits our chosen classification problem. 

All the systems' specifications and the hyperparameters can be found in table~\ref{tab:hyperparams}.

%============================================================================================
% Hyperparameters
\begin{table}[htpb]
\caption{Hyperparameters and System Specifications}
    \label{tab:hyperparams}
    \centering
    \resizebox{0.85\columnwidth}{!}{
    \setlength{\tabcolsep}{2pt}
    \begin{tabular}{|l|c|c|}
    \hline
    \textbf{System}& \textbf{Hyperparameter} & \textbf{Value} \\ \hline
    \textbf{Common}& Optimiser & Adam \\ \cline{2-3}
                   & lr & 1e-4 \\ \cline{2-3}
                   & lr-scheduler & \makecell{step\\step size: 10\\ $\gamma$: 0.75} \\ \cline{2-3}
                   & Loss function & Binary cross-entropy \\ \cline{2-3}
                   & Batch size & 64\\ \cline{2-3}
                   & Epochs & 100 \\ \hline
    \textbf{Quantum} & Framework & Pennylane \\ \cline{2-3}
                    & Ideal simulator & \makecell{AER(Qiskit)\\default.qubit(Pennylane)} \\ \cline{2-3}
                    & Noisy simulator & \makecell{Forte-1(IonQ)\\$r_{1_q}:2.67e-4$\\$r_{2_q}:4.94e-3$} \\ \cline{2-3}
                    & Number of Qubits & [3 to 10] \\ \cline{2-3}
    \hline
    \bottomrule
    \end{tabular}
    }
   
\end{table}
%============================================================================================

\section{Results}\label{sec: results}
In this section, we present the experimental results of our hypothesis, \ie, quantum transfer learning can be successfully used to enhance the performance of a light, suboptimal classical model. 

\subsection{Classical Models}
\subsubsection{\textbf{Baseline:}}
We used a small classical convolutional model as the baseline. We assume that this small, lightweight model suits a resource-constrained environment. The baseline suffers from high bias and produces a moderate accuracy of 73\%. From the table~\ref{tab:performance_comp}, we can see that this model's precision, recall, F1 score, and AUC are 0.69, 0.74, 0.71, and 0.83, respectively. A poor recall in healthcare applications is considered bad, increasing the chances of a false negative. The baseline, as it produces a poor recall of 0.74, has a high chance of missing a positive case, eventually failing to diagnose and address a demented patient.

\subsubsection{\textbf{Classical Fine Tuning \textit{via.} transfer learning}:}
We fine-tuned the classical dense layers of the baseline by freezing the initial convolutional layers at their pretrained weights and retraining the dense layers for another \textit{100} epochs after initialization. This method improved the performance quite a bit as expected. With classical fine-tuning, the test accuracy improved by 18\%. Performance improved along all the other matrices by roughly 20\%, with a 13.5\% improvement in AUC. The classical transfer learning based model produced a recall of 0.87, which is an 18\% improvement over the baseline.

\subsection{Quantum Transfer Learning}
We applied quantum transfer learning by replacing the last set of dense layers with a \textit{dressed quantum network} (DQN), freezing the convolutional layers of the pretrained baseline and training the DQN over \textit{100} epochs. 

The architecture of the DQN varies with the number of qubits and the number of repetitions used in the quantum circuit. We performed a grid search over 3 to 10 qubits and 2 to 4 repetitions of the ansatz (Figure~\ref{fig:ansatzA}). Based on the test accuracies recorded in the ideal simulation, we chose the 6-qubits and 4-repetitions setup as the best-performing model. The test accuracy of this setup was recorded as 90.50\%, which was the highest among the 24 total resulting configurations. 

We also faced the barren plateau problem~\cite{mcclean2018barren} while training the QTL-based models. As we saw an almost equivalent performance across different VQC setups, signifying that the optimisation landscape was quite flat, and therefore it was challenging for the cost to converge to a global minimum. 

Then we simulated the 6-qubits, 4-repetitions setup in the Forte-1 noisy environment. We compared the performance of this best-performing QTL setup in the noisy environment with the baseline and the most efficient classical transfer learning-based model in table~\ref{tab:performance_comp}. Here, we observed that the QTL produced 91.29\% test accuracy, which is a 25\% improvement over the baseline. We also achieved a roughly 27\% improvement in precision, recall, and F1 score. With 94\% recall, the QTL-based approach delivered the most reliable model in our resource-constrained environment. The AUC score was also improved by 11\% to 0.92, signifying that the quantum transfer learning-based model is better at distinguishing between the positive and negative samples.

We also observed improvements with the QTL-based hybrid model compared to the classically fine-tuned model (CTL). The test accuracy, precision, recall, and F1 score were almost 5 to 7\% higher than CTL. However, we saw a nominal decrease of 1.8\% in the AUC score. That means the classical transfer learning produced a model slightly better at distinguishing between the positive and negative data compared to the quantum transfer learning. But the latter has less chance of missing out a true positive due to a higher recall score of 0.94. 

%============================================================================================
%Performances of the Baseline, Classically Fine-tuned and QTL-based Approaches and their Quantitative Comparison

\begin{table}[htpb]
\caption{Performances of the Baseline, Classically Fine-tuned, and QTL-based Approaches and their Quantitative Comparison}
    \label{tab:performance_comp}
    \centering
    \resizebox{\columnwidth}{!}{
    \setlength{\tabcolsep}{2.5pt}
    \begin{tabular}{|l|c|cc|ccc|}
        \hline
        \textbf{Models--$\blacktriangleright$}& \textbf{Baseline}& \textbf{CTL}&  & \textbf{QTL}& &  \\ \cline{3-7}

        & & Orig. & \textbf{\makecell{$\triangle$Imp.\\Base(\%)}} & Orig. & \textbf{\makecell{$\triangle$Imp.\\Base(\%)}} & \textbf{\makecell{$\triangle$Imp.\\CTL(\%)}} \\
        \hline
        \textbf{Test Acc (\%)} & 73.10&	86.36&	18.14&	91.29&	24.88& 5.71\\ \hline
        \textbf{Precision} & 0.6875&	0.8320&	21.02&	0.8750&	27.27& 5.17\\ \hline
        \textbf{Recall} & 0.7394&	0.8739&	18.19&	0.9412&	\textbf{27.29}&  \textbf{7.7}\\ \hline
        \textbf{F1 score} & 0.7125&	0.8524&	19.64&	0.9069&	27.28& 6.39\\ \hline
        \textbf{AUC} & 0.8282&	0.9399&	13.49&	0.9232&	11.47&  -1.78\\ \hline
        
    \bottomrule
    \end{tabular}
    }
    \footnotesize{\\Reportedly, the QTL-based model used an ansatz with six qubits and four repetitions. Improvements are shown next to the original performance scores of individual models, derived from the baseline.}
\end{table}
%============================================================================================

\section{Conclusion and Future Work}\label{sec: conclusion}
This work demonstrates an application of quantum transfer learning to detect dementia. We formulated a binary classification problem based on the OASIS 2 MRI dataset. We begin the experiment by training a classical baseline model, then enhance its performance using several configurations of quantum transfer learning. Starting with the weak classical model, we show that quantum transfer learning enhanced its performance in predicting dementia. This approach shows that, in several cases, we may only access a weak classical model. However, QTL can be one possible solution to improve it instead of discarding it entirely. We have also shown that the hybrid quantum models can be very resilient in noisy environments without deploying rigorous error mitigation techniques. However, the noise impact will rise with the number of qubits in use, adding a crucial constraint to the solution. 

In this experiment, we performed a grid search along the number of qubits (from 3 to 10) and the number of repetitions (from 2 to 4). Based on the time and hardware constraints, the search can be extended across the number of qubits, repetitions, and the design of the ansatz. We can also try different strategies~\cite{sack2022avoiding} to avoid the barren plateau problem altogether.

Although quantum machine learning and quantum computation are still in their infancy, with the evolution of more robust and noiseless quantum processors, they will be able to address many real-world biomedical problems, like predicting dementia, in the future.

\begin{acks}
    This research used the resources of the Oak Ridge Leadership Computing Facility, which is a DOE Office of Science User Facility supported under Contract DE-AC05-00OR22725. This research was also supported by the U.S. Department of Energy (DOE) under Contract No. DE-AC02-05CH11231, through the Office of Science, Office of Advanced Scientific Computing Research (ASCR), Exploratory Research for Extreme-Scale Science and Accelerated Research in Quantum Computing.
\end{acks}

\bibliographystyle{ACM-Reference-Format}
\bibliography{QTL_dem}

\balance
%\vspace{12pt}

\end{document}